\documentclass[a4paper,12pt]{article}
\usepackage{jheppub}
\usepackage{physics}
\usepackage{amsmath}
\usepackage{amssymb}
\usepackage{amsfonts}
\usepackage{xcolor}
\usepackage[backgroundcolor=pink, textsize=scriptsize]{todonotes}
\usepackage{lmodern}
\usepackage[utf8]{inputenc}
\usepackage[T1]{fontenc}
\usepackage{caption}
\usepackage{subcaption}
\usepackage{chngcntr}
\counterwithout{equation}{section}
\definecolor{refkey}{RGB}{0,255,0}
\definecolor{labelkey}{RGB}{255,0,0}
\hypersetup{
  colorlinks=true,
  citecolor=magenta,
  linkcolor=blue,
  urlcolor=violet
 }

\newcommand{\lpl}{\ell_{\mathrm{Pl}}}
\newcommand{\GN}{G_{\mathrm{N}}}
\newcommand{\Lads}{L_{\mathrm{AdS}}}
\newcommand{\wq}{\mathcal{W}_Q}

\newcommand{\mi}{\mathrm{i}}
\newcommand{\baa}{\begin{equation}\begin{aligned}}
\newcommand{\ea}{\end{aligned}\end{equation}}
\newcommand{\be}{\begin{equation}}
\newcommand{\ee}{\end{equation}}
\newcommand{\wt}{\widetilde}

\title{\center{ \fontfamily{lmr}\selectfont \boldmath Large Quantum Gravity Fluctuations
of BTZ Black Holes}}

\author[a,b]{ \center \fontfamily{lmr}\selectfont Ben Freivogel}
\author{and}
\author[a]{Upamanyu Moitra}

\affiliation[a]{\begin{center}
    Institute for Theoretical Physics, Institute of Physics, Universiteit van Amsterdam,
Science Park 904, 1098 XH Amsterdam, The Netherlands
\end{center}}

\affiliation[b]{\begin{center}
    GRAPPA, Universiteit van Amsterdam,\\
Science Park 904, 1098 XH Amsterdam, The Netherlands

\href{mailto: b.w.freivogel@uva.nl}{\textup{\texttt{ b.w.freivogel@uva.nl}}}\textup{,}
\href{mailto:u.moitra@uva.nl}{\textup{\texttt{u.moitra@uva.nl}}}

\end{center}}

\abstract{
We study the quantum fluctuations of the black hole horizon in three-dimensional Anti-de Sitter (AdS) spacetime. We define a precise protocol to calculate the horizon fluctuations and define a corresponding ``quantum width'' of the horizon. We relate the horizon fluctuations to boundary correlation functions via holography. Working in perturbative quantum gravity, we find that the quantum width is typically of order $(G_\mathrm{N} L_{\mathrm{AdS}}^3 )^{1/4}$, which is parametrically larger than the Planck scale.  In detail, the quantum width depends on the scale at which it is measured, diverging logarithmically in the UV.  Our results give the most rigorous evidence to date of gauge-invariant fluctuations at scales much larger than the Planck scale within perturbative quantum gravity.
}

\makeatletter
\gdef\@fpheader{{}}
\makeatother
\begin{document}
\maketitle
\flushbottom

\section{Introduction}\label{sec:intro}

One of the most intriguing questions in physics concerns the possibility of observing effects of quantum gravity. Such effects are usually believed to be of Planck scale ($\sim 10^{-33}$ cm) and hence unobservably small. There have been suggestions \cite{Verlinde:2019xfb, Verlinde:2019ade} that such effects can be non-perturbatively enhanced in a parametrically large way. Perturbation theory, on the other hand, seems to support the general belief that such effects are very small.

A possible situation where perturbative fluctuations may be large are the quantum fluctuations of black hole horizons. Remarkably, this question does not seem to have been answered in general, perhaps because it is not obvious how to define the fluctuations of the horizon in a gauge-invariant way. Previous work has focused mainly on the fluctuations in the intrinsic geometry of the event horizon. Here, we want to define and calculate fluctuations in the location of the event horizon. 

These fluctuations were estimated by Marolf \cite{Marolf:2003bb} in 2003, and dubbed the `quantum width' of the black hole horizon; more recent work includes \cite{Parikh:2024zmu}. Marolf's estimate for the quantum width $\wq$ of Schwarzschild black holes in $D$ spacetime dimensions is
\begin{equation}
    \wq^D \sim  \GN r_s^2,
    \end{equation}
so that the quantum width is a sort of geometric mean between the Planck scale and the Schwarzschild radius. 

Here, we want to perform a more careful calculation of these fluctuations. As a first step, we will need to define the quantum width in a gauge-invariant way, which has not been done in previous work. Then we will calculate the fluctuations in ordinary perturbative quantum gravity to leading order.  
In this paper, we focus on three dimensions with a negative cosmological constant and calculate quantum fluctuations of the horizon of the Ba\~{n}ados-Teitelboim-Zanelli (BTZ) black hole \cite{Banados:1992wn}.

Our results will also resolve two puzzling features of the Marolf estimate. First, we expect the fluctuations to depend on the length scale at which the horizon is probed, not only on the Schwarzschild radius. Second,
 the Marolf result agrees with the thermodynamic estimate ${\delta r \over r_s} \sim {1 \over \sqrt{S_{BH}}}$ only in 4 spacetime dimensions \cite{Verlinde:2019xfb, jindong}

\paragraph{Summary of Results.} For the impatient reader, we summarize our main results. The simplest summary is that the quantum width is typically of order
\begin{equation}
\boxed{
    \wq^4 \sim \GN \Lads^3.
    }
\end{equation}
However, the more precise results do depend on the resolution of the measurement. The quantum fluctuation of the horizon at a precisely defined point diverge, as usual in QFT. Therefore, to get a finite answer, we need to smear the observable in space and/or time. It is worth noting that this length scale was also previously observed in \cite{Chen:2017dnl} in a different context.

Typically, in QFT, the fluctuations of observables depend on the spatial as well as time resolution of the measurement. However, in 3 dimensions there are only `boundary gravitons', which are either left- or right-moving. Therefore, smearing in space is equivalent to smearing in time, so the result depends only on one smearing length, at least in our leading order calculation.

Considering a large AdS black hole, $r_0 \gg \Lads$, the quantum fluctuations of a piece of the horizon of proper size $\hat \sigma$ are given by
\baa
     \expval{\wq^4 } = \begin{cases}
        64 \pi^2 \GN \Lads^3  \log { \Lads \over \hat \sigma}  + \cdots, &\quad \hat \sigma  \ll \Lads, \\
        \dfrac{16\pi^{-1/2} \GN \Lads^4 }{\hat \sigma} + \cdots, &\quad \hat \sigma  \gg \Lads. 
    \end{cases}
\ea
The logarithmic divergence for small regions comes from the UV divergences mentioned above. For large distances, the fluctuations are suppressed. This is consistent with the lore that  AdS-sized regions of the horizon fluctuate independently of each other, so that the power law suppression comes from $1/\sqrt{N}$ statistics, as we describe in more detail below.

It is also natural to Fourier transform on the horizon circle. The correlator of a mode with angular momentum $m$ is UV-finite and has a simple time dependence,
\baa
\expval{ (\wq^2)_m (t) (\wq^2)_n (t') } = 32 \pi^2 \GN \Lads^3  \delta_{m +n, 0} \bqty{ e^{ \mi \frac{m}{\Lads} (t-t') } \widetilde{\mathcal{S}}_{-m} + e^{ -\mi \frac{m}{\Lads} (t-t') } \widetilde{\mathcal{S}}_{m} },\label{delamdisccorr11}
\ea
where
\baa
\widetilde{\mathcal{S}}_{m}  = \frac{m}{\qty(m^2 + 4\pi^2 \Lads^2 T^2) \qty( 1 - \exp( - \frac{m}{\Lads T}  ) ) }.  \label{tildesmdef11}
\ea
The fact that $\wq^2$ instead of $\wq$ appears naturally in the correlator comes from the definition of the quantum width, which we give now.

\paragraph{Definition of the Quantum Width.} We want to define quantum fluctuations in the location of the black hole horizon in a gauge-invariant way. The following definition is one natural choice: 
\begin{enumerate}
    \item  Let an observer fall freely from infinity, beginning at rest in the frame selected by the black hole. 
    \item In the presence of perturbations, the proper time for the observer to cross the event horizon fluctuates, $\tau = \tau_c + \Delta \tau$, with $\tau_c$ the classical time to reach the event horizon.
\end{enumerate}
This already defines a gauge-invariant observable, and we can calculate $\expval{(\Delta \tau)^2}$ in perturbation theory. But to connect to previous literature and build intuition, we want to convert to a proper length, which has been called the `quantum width' of the black hole horizon.

The quantum width is supposed to capture the width of the region where it is uncertain whether a signal sent out will reach infinity, or fall into the black hole. An infalling observer who wants to reliably send a signal to infinity  must send it at a proper time before $\tau_c - \sqrt{\expval{(\Delta \tau)^2}}$. We can calculate the typical distance of the observer from the event horizon by calculating, in the background geometry, the proper distance from the horizon to the observer's location at this time. This defines the quantum width, which in this paper we call $\wq$. 

\begin{figure}
    \centering
    \includegraphics[width=0.6\linewidth]{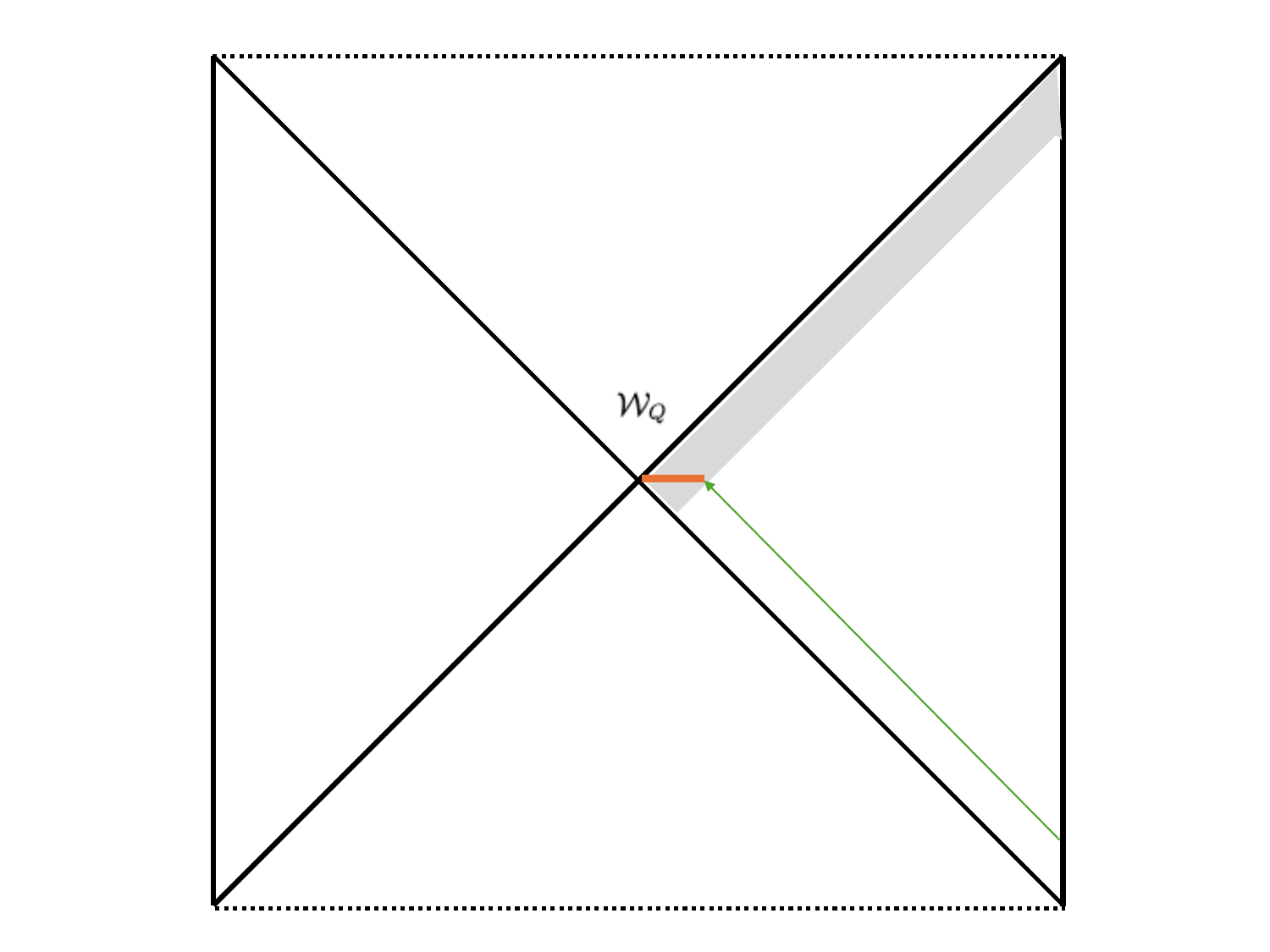}
    \caption{A schematic depiction of how we calculate the quantum width $\wq$ as a proper distance from the unperturbed horizon to a point on a null geodesic (green). The gray region represents the quantum fluctuations of the horizon.}
    \label{fig:qw}
\end{figure}

We actually calculate a slight variation on this observable:  we work with null geodesics instead of timelike geodesics for simplicity --- so instead of discussing the proper time fluctuations, we calculate fluctuations in the affine parameter $\lambda$. The set-up for our calculation is schematically depicted in Figure \ref{fig:qw}. We expect that a calculation using timelike geodesics would yield the same quantum width, up to possible order-one factors.

Let us guide the reader through the contents of this article. In \S\ref{sec:set-up}, we establish the basic set-up and make a calculation of horizon fluctuations with thermodynamic considerations. In \S\ref{sec-corrf}, we study these fluctuations and the quantum width in detail with holographic correlation functions.  In \S\ref{sec-various}, we extend the discussion to a more general set-up with a finite horizon size. In \S\ref{sec:limits}, we discuss some physically interesting limits. We conclude the paper in \S\ref{sec-discuss} with a discussion of another possible observational protocol which could potentially detect such fluctuations more generally and some general comments including future prospects.

\section{Set-Up and Thermodynamic Considerations}\label{sec:set-up}

We study pure Einstein gravity in three dimensions with a negative cosmological constant. The action is given by
\begin{equation}
    I = \frac{1}{16\pi \GN} \int \dd[3]{x} \sqrt{-g} \pqty{ R + \frac{2}{\Lads^2} } + I_{\mathrm{GHY}} + I_{\mathrm{ct}}\label{ehac},
\end{equation}
where $\Lads$ is the radius of the Anti-de Sitter (AdS) spacetime. In three dimensions, the Newton constant $\GN$ is equivalent to the Planck length
\begin{equation}
G = \lpl. \label{gdef}
\end{equation}
Since several length scales will appear in this article, we are going to explicitly retain factors of $\GN$ and $\Lads$. 

Let us consider the simplest possible set-up for our problem with a non-rotating BTZ black hole \cite{Banados:1992wn}, with the metric
\begin{equation}
    \dd{s}^2 = - f(r) \dd{t}^2 + \frac{\dd{r}^2}{f(r)} + r^2 \dd{\phi}^2, \label{nrbtz}
\end{equation}
where the red-shift factor is given by
\begin{equation}
    f(r) = \frac{r^2}{\Lads^2} - 8 \GN M = \frac{r^2 - r_0^2}{\Lads^2},\label{rsfbtz1}
\end{equation}
where $M$ is the mass of the black hole.
The horizon radius is given by
\begin{equation}
    r_0 = \sqrt{8 \GN M \Lads^2}. \label{hrad1}
\end{equation}
All the thermodynamic quantities are simple functions of the mass or equivalently the horizon radius. The Hawking temperature is 
\begin{equation}
    \beta^{-1} = T = \frac{r_0}{2\pi \Lads^2} = \frac{1}{2\pi} \sqrt{\frac{8\GN M}{\Lads^2}}. \label{temp1}
\end{equation}
On the other hand, the Bekenstein-Hawking entropy is given by
\begin{equation}
    S = \frac{\pi r_0}{2\GN} = \pi \sqrt{\frac{2M\Lads^2}{\GN}}. \label{ent1}
\end{equation}

It will sometimes be convenient to trade the mass $M$ for the inverse temperature $\beta$ by inverting the relation \eqref{temp1},
\begin{equation}
    M = \frac{\pi^2 \Lads^2}{2  \GN} \frac{1}{\beta^2}, \label{mbtrel}
\end{equation}
which yields for the entropy
\begin{equation}
    S = \frac{\pi^2 \Lads^2}{\GN} \frac{1}{\beta} = 2 M \beta, \label{entbt}
\end{equation}
and free energy
\begin{equation}
    F = -\frac{\pi^2 \Lads^2}{2\GN} \frac{1}{\beta^2} =  - \frac{S}{2 \beta}. \label{febet}
\end{equation}

It is worth highlighting that while there does not exist an analog of ``small'' black holes in $\mathrm{AdS}_3$, there is nevertheless a Hawking-Page transition \cite{Hawking:1982dh}. The thermal global AdS state carries zero entropy to the leading order in $\GN$, but has a negative energy given by $M = -1/(8\GN)$, leading to the free energy
\baa
F_{\rm thermal} = - \frac{1}{8 \GN}. \label{ftherm}
\ea
A comparison with the black hole free energy \eqref{febet}  thus shows that the canonical ensemble is dominated by black holes above a critical temperature $T > T_{\rm crit}$ given by
\baa
T_{\rm crit} = \frac{1}{2\pi \Lads}. \label{tcrit}
\ea
Below this temperature, thermal AdS is the dominant configuration.

\subsection{Thermodynamic Calculation of Horizon Fluctuations}
We want to calculate the fluctuations of the horizon radius in the canonical ensemble. While doing so,  we keep $\GN$ and $\Lads$ fixed. From eq. \eqref{hrad1}, we immediately see that
\begin{equation}
    \frac{\Delta r_0}{r_0} = \frac{1}{2} \frac{\Delta M}{M}. \label{rmrel}
\end{equation}
Let us now establish the following important relation for energy fluctuations in the canonical ensemble,
\begin{equation}
    \frac{\Delta M}{M} = \frac{2}{\sqrt{S}}. \label{mflu1}
\end{equation}

We have the spread of the energy,
\begin{equation}
    \Delta M = \sqrt{\expval{M^2} - \expval{M}^2}. \label{deldef}
\end{equation}
The variance  $(\Delta M)^2$ of the energy is most easily obtained from the canonical partition function $Z(\beta)$,
\begin{equation}
    \expval{M^2} - \expval{M}^2 = \pdv[2]{\beta} \log Z(\beta) = -\pdv[2]{\beta} (\beta F) ,  \label{mflu2}
\end{equation}
where the partition function is defined in the usual way, $Z(\beta) \equiv \sum_M \exp(-\beta M)$. We find using eq. \eqref{febet},
\begin{equation}
    \expval{M^2} - \expval{M}^2  = \frac{\pi^2 \Lads^2}{\beta^3 \GN} = \frac{S}{\beta^2}. \label{mflu3}
\end{equation}
 Using this relation and eq. \eqref{entbt}, we recover the desired relation \eqref{mflu1}. Therefore, we obtain from \eqref{rmrel},
\begin{equation}
    \frac{\Delta r_0}{r_0} = \frac{1}{\sqrt{S}} ,\label{candelrsqflu}
\end{equation}
\begin{equation}
    (\Delta r_0)^2 = \frac{2}{\pi} \GN \, r_0. \label{candelrsq}
\end{equation}
Thus, in this case, the fluctuation of the horizon radius is the geometric mean of the Planck length $\lpl$ and the horizon size $r_0$, which is also seen in higher dimensions. Quite reassuringly, it scales with $r_0$. Note that the variance \eqref{candelrsq} does not depend on the AdS radius $\Lads$.

\subsection[The Quantum Width $\wq$ in the Canonical Ensemble]{\boldmath The Quantum Width $\wq$ in the Canonical Ensemble}\label{subsec-qw}

We are now interested in an invariant characterization of the horizon fluctuations in terms of some observable. One possible candidate is the affine parameter along an infalling null geodesic. For simplicity, let us consider a radially infalling ray for which we have
\baa
    - f(r) \pqty{ \dv{t}{\lambda} }^2 + \frac{1}{f(r)} \pqty{ \dv{r}{\lambda}}^2 = 0. \label{ngeo1}
\ea
The quantity $f(r) \dd{t}/\dd{\lambda}$ is conserved --- by an appropriate scaling of $\lambda$, we can choose this conserved quantity to be unity. Thus, if the null ray starts from the boundary --- for infra-red (IR) regulation purposes, we take it to be some fixed value of the radial coordinate $r=r_C$ far from the black hole --- then we have the affine parameter at the instant of reaching the black hole horizon given by
\baa
    \lambda = r_C - r_0\label{aff1}.
\ea
If we keep the cut-off surface at $r = r_C$ fixed and consider fluctuations of the location of the black hole horizon, eq. \eqref{aff1} tells us that the fluctuation of the affine parameter is given by
\baa
    \Delta \lambda = \Delta r_0. \label{affdel2}
\ea
which is independent of the IR regulator. Note that this quantity is well-defined when we send the IR regulator $r_C$ to infinity.

As discussed in the introduction,  
$\wq$ is defined as (we restrict ourselves to the near-horizon region)
\baa
    \wq = \int_{r_0}^{r_0 + \Delta r_0} \frac{\dd{r}}{\sqrt{f(r)}} \approx \frac{2 \sqrt{\Delta r_0}}{\sqrt{f'(r_0)}}. \label{defqw1}
\ea
Using the relation between the temperature and the derivative of the emblackening factor as in eq. \eqref{temp1},
\begin{equation}
    4 \pi T = f'(r_0)
\end{equation}
the relation between the quantum width and the fluctuation in $r$ is conveniently packaged as
\begin{equation}
    \wq^2= {\Delta r_0 \over \pi T}.
    \label{widthfromdeltar}
\end{equation}
 Due to the quadratic nature of this formula, the natural quantities that appear in perturbation theory are correlators of $\Delta r_0$, which become correlators of $\wq^2$.

For our case of BTZ black holes, the canonical ensemble gives
\baa
    \wq^2 = 2 \sqrt{\frac{2}{\pi} } \Lads^2 \sqrt{\frac{\GN}{r_0}}. \label{qwcase1}
\ea
We find that the width depends explicitly on the AdS length scale $\Lads$ and, somewhat surprisingly, also that $\wq$ goes down as $r_0$ increases.

An intuitive physical picture of the scenario is as follows. For a black hole of radius much bigger than the AdS length scale $r_0 \gg \Lads$, we can imagine the black hole circumference (also the area in this case) is divided into cells with a proper size $\Lads$,
\baa
    N_{\rm cell} \sim \frac{r_0}{\Lads} \label{ncelldef}.
\ea
For a fluctuation of the radius in each cell, the average fluctuation is given by 
\baa
    \Delta r_{\rm avg} \sim {1 \over N_{\rm cell}} \sum_{\rm cells} \Delta r_{\rm cell}. \label{dertot}
\ea
The lore of large AdS black holes
tells us that each AdS-sized cell behaves roughly independently. Therefore, we take the fluctuations of the cells to be independent and identically distributed, yielding
\baa
    \pqty{\Delta r_{\rm avg}}^2 \sim {1 \over N_{\rm cell}} \pqty{\Delta r_{\rm cell} }^2. \label{drtotsq}
\ea

Now, we want to assume that the fluctuations of each cell are independent of the total size $r_0$ for large $r_0$. The right assumption is that the quantum width for each cell is independent of $r_0$. Therefore, we get
\begin{equation}
   ( \wq^2)_{\rm avg} \sim \sqrt{1 \over N_{\rm cell}} (\wq^2)_{\rm cell} \sim \sqrt{L_{\rm AdS} \over r_0}(\wq^2)_{\rm cell}.
\end{equation}
This reproduces the $r_0$ dependence seen in \eqref{qwcase1}. Comparing the two formulas implies that the quantum width, when measured on AdS scale is 
\begin{equation}
    \wq^{\rm cell} \sim (\GN L_{\rm AdS}^3)^{1/4}\ . \label{newscale}
\end{equation}
We will see this length scale emerge repeatedly in our more careful analysis below.

\section{Horizon Fluctuations from Holographic Correlators}\label{sec-corrf}

The thermodynamic arguments in the previous section, while being completely general in nature, do not give a reliable calculation of a gauge-invariant observable. In this section, we define a physically motivated observable and study its correlation functions. We consider small quantum fluctuations around the BTZ black hole geometry. At leading order, fluctuations in the geometry do not mix with other fields, so these perturbed spacetimes are Ba\~nados geometries \cite{Banados:1998gg}, which are characterized by a pair of independent functions $\ell^\pm (x^\pm)$. 

We proceed in two steps. First, we calculate the affine parameter to the horizon in a given Ba\~nados geometry. Then, to calculate the fluctuations, we sum over such geometries with a weighting given by perturbative quantum gravity. It is convenient to do this by relating correlators of the gravitational perturbations to stress tensor correlators in the holographic dual.

\subsection{Near-BTZ Geometry}\label{subsec-nearbtz}

The Ba\~nados geometry \cite{Banados:1998gg}, after an appropriate analytic continuation to Lorentzian signature, is given by
\baa
     \dd{s}^2 = \Lads^2 \frac{\dd{z}^2}{z^2}  -\pqty{\frac{\Lads^2}{z^2} + \frac{z^2}{\Lads^2} \ell_+ (x^+) \ell_- (x^-)   }\dd{x^+} \dd{x^-} + \pqty{ \ell_+ \dd{x^+}^2 + \ell_- \dd{x^-}^2 }, \label{banad3}
\ea
where 
\baa
x^\pm = t \pm \Lads \phi
\label{xpmdef}
\ea
are the boundary light-cone coordinates. We will sometimes take $\phi$ to be a non-compact coordinate, in which case, strictly speaking, there is no black hole. The black hole is present when there is a periodic identification of $\phi$,
\baa
    \phi \sim \phi + 2\pi \label{phiperd}.
\ea
However, even for a non-compact $\phi$, the geometry can be interpreted as having a Rindler horizon. We will see that the most important features of the calculation go through for the Rindler horizon.  Our main conclusions, therefore, are of greater validity involving general horizons.

We want the background geometry to be that of the the static BTZ black hole discussed in \S\ref{sec:set-up}, for which we have
\baa
    \ell^{(0)}_\pm = 2 \GN M = \frac{r_0^2}{4\Lads^2}. \label{ellonr}
\ea
Under the coordinate transformation
\baa
    z = \frac{2\Lads^2}{r_0^2} \pqty{r - \sqrt{r^2 - r_0^2} }, \label{zrct}
\ea
we recover the non-rotating BTZ metric in the standard radial coordinate, eq. \eqref{nrbtz}.

Since the Ba\~nados geometry is locally ${\rm AdS}_3$, it is possible to locally gauge away the functions $\ell_\pm$ by a diffeomorphism. The exact non-linear diffeomorphism is known in the literature \cite{Roberts:2012aq}. If we start with the metric in the Poincar\'{e} patch,
\baa
    \dd s^2 = \frac{\Lads^2}{u^2} \pqty{ \dd{u}^2 - \dd{y}^+ \dd{y}^- },  \label{initpoinc}
\ea
then the following diffeomorphism takes us to the metric \eqref{banad3},

\baa
        y^\pm &= f_\pm (x^\pm) + \frac{2z^2 f'_\pm(x^\pm)^2 f''_\mp (x_\mp) }{4f'_+ (x^+) f'_- (x^-) - z^2 f''_+ (x^+) f''_- (x^-) }, \\
        u &= z \frac{4 \pqty{f'_+ (x^+) f'_- (x^-)}^{3/2} }{4f'_+ (x^+) f'_- (x^-) - z^2 f''_+ (x^+) f''_- .(x^-)},
\ea
with
\baa
    \ell_\pm (x^\pm)  = -\frac{1}{2} \Lads^2 \qty{ f_\pm (x^
    \pm) , x^\pm },  \label{nonlinlchange}
\ea
where $\{ f(z), z \}$ refers to the Schwarzian derivative, defined as
\baa
    \qty{ f (z) , z } \equiv  \frac{f'''(z)}{f'(z)} - \frac{3}{2} \pqty{ \frac{f''(z)}{f'(z)} }^2. \label{defschwarzian}
\ea

Our present goal is to  study small perturbations about the BTZ background.  We are therefore interested in diffeomorphisms that take us from one Ba\~nados metric to another nearby one. This is achieved by the following infinitesimal diffeomorphism \cite{Roberts:2012aq},
\baa
    x^\pm &\to x^\pm + \epsilon \pqty{ f_\pm (x^\pm) + \frac{\Lads^2 z^4 \ell_\mp (x^\mp)  f''_\pm (x^\pm) + \Lads^4 z^2 f''_{\mp} (x^\mp) }{2 \pqty{ \Lads^4 - z^4 \ell_+(x^+) \ell_- (x^-) } } }, \\
    z &\to z \pqty{1 + \frac{1}{2} \epsilon \qty(f'_+(x^+) + f'_-(x^-)) },
    \label{lindiffeo1}
\ea
where $\epsilon \ll 1$. Under this transformation, a different Ba\~nados metric is obtained with a new pair of transformed $\ell_\pm (x^\pm)$ given by
\baa
    \ell_\pm (x^\pm) \to \ell_\pm (x^\pm) + \epsilon \pqty{2 \ell_\pm (x^\pm) f'_\pm (x^\pm) +  \ell'_\pm (x^\pm) f_\pm (x^\pm) - \frac{1}{2} \Lads^2 f'''_\pm (x^\pm)   }  .  \label{linshiftelles}
\ea

\subsection{Null Probes in the Fluctuating Geometry}\label{subsec-npfg}

\begin{figure}
\centering
\begin{subfigure}{.33\textwidth}
    \centering
    \includegraphics[width=\linewidth, page =1]{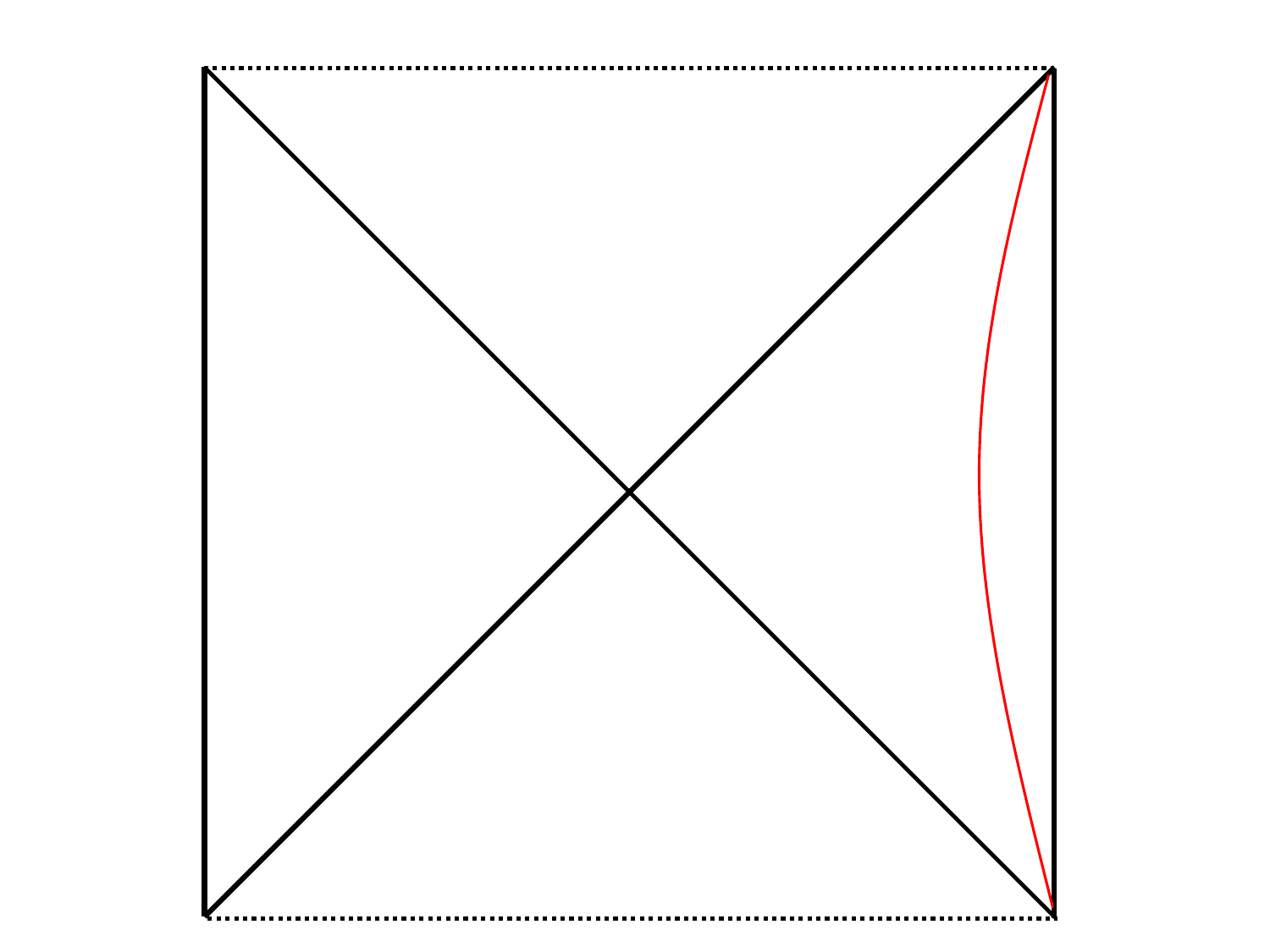}
    \caption{}
    \label{fig:cpda}
\end{subfigure}%
\begin{subfigure}{.33\textwidth}
    \centering
    \includegraphics[width=\linewidth, page = 2]{Causal_Diagrams_Final.pdf}
    \caption{}
    \label{fig:cpdb}
\end{subfigure}%
\begin{subfigure}{.33\textwidth}
    \centering
     \includegraphics[width=\linewidth, page =3]{Causal_Diagrams_Final.pdf}
    \caption{}
    \label{fig:cpdc}
\end{subfigure}
\caption{The sequence of operations for obtaining a perturbed geometry: In (a) is the unperturbed BTZ black hole with the metric \eqref{nrbtz} and a boundary surface at $z = \delta$. The perturbed geometry given by \eqref{pertban3} is depicted in (b). The inverse diffeomorphism is performed to return to the unperturbed bulk in (a), now at the cost of having a ``wiggly'' boundary surface given by $z = \delta\bqty{1 + \epsilon g(x^+, x^-)}$, which is shown in (c).}
\label{fig:cpd}
\end{figure}

We would now like to generalize the procedure described in \S\ref{subsec-qw} where we related the fluctuation of the black hole horizon to the fluctuation of the affine time taken by a null ray sent radially inward towards the black hole horizon from the asymptotic boundary. As before, we need an IR regularization of the boundary cut-off surface. The boundary surface in the previous discussion was placed at a constant value of the radial coordinate, $r=r_C$. In the $z$-coordinate, this would translate to a constant-$z$ surface by eq. \eqref{zrct}, say $z = \delta$, where $\delta$ is related to $r_C$ by the same equation. This unperturbed surface in the unperturbed geometry is depicted in Figure \ref{fig:cpda}.

It is important to clarify our protocol for obtaining a perturbed black hole geometry even though the bulk diffeomorphism relates one geometry to the other. Note that our protocol for sending in the null ray depends both on the bulk geometry and the boundary conditions satisfied by it. Starting with the aforementioned unperturbed geometry with a constant-$z$ boundary (Figure \ref{fig:cpda}), we apply the infinitesimal diffeomorphism to obtain a different bulk geometry, but apply the \emph{same} boundary condition as before, as shown in Figure \ref{fig:cpdb}. We send a null ray from the boundary of this geometry. Since we change the bulk, but do not change the boundary, in a way made precise below, the results are different. To simplify the calculation, it is now convenient to perform the inverse diffeomorphism on \emph{both} the bulk and the boundary so that the bulk looks like the unperturbed BTZ geometry but the boundary is no longer at a constant $z$, but wiggly, as shown in Figure \ref{fig:cpdc}. In this geometry, the calculation involving the null geodesic is simple. Note that the situations described in Figures \ref{fig:cpdb} and \ref{fig:cpdc} are physically equivalent --- only a different coordinate system is being used to facilitate the calculation. 

The unperturbed background geometry with the static BTZ black hole is simply the metric \eqref{banad3} with $\ell_+ = \ell_- = \ell^{(0)} = 2 \GN M = r_0^2 / 4 \Lads^2$ (see eq.  \eqref{ellonr}).  We can normalize the affine parameter as in eq. \eqref{aff1} in this coordinate system giving,
\baa
\lambda = - \pqty{ \frac{\Lads^2}{z} + \ell^{(0)}z  }  + \pqty{ \frac{\Lads^2}{\delta} + \ell^{(0)} \delta  }.  \label{affinezee}
\ea
This affine parameter satisfies the initial conditions which define the cut-off surface and the normalization
\baa
\eval{\lambda}_{z= \delta} = 0, \qquad  \eval{\dv{\lambda}{z}}_{z = \delta} = \frac{\Lads^2}{\delta^2} - \ell^{(0)}.  \label{affineinit}
\ea

We now perform the infinitesimal diffeomorphism defined by \eqref{lindiffeo1}. Let us define the coordinate system precisely.  We go from the coordinates $(z,  x^\pm)$ to $(\tilde{z}, \tilde{x}^\pm)$ by the same relation,
\baa
        x^\pm &= \tilde{x}^\pm + \epsilon \pqty{ f_\pm (\tilde{x}^\pm) + \frac{\Lads^2 \tilde{z}^4 \ell^{(0)}  f''_\pm (\tilde{x}^\pm) - \Lads^4 z^2 f''_{\mp} (\tilde{x}^\mp) }{2 \pqty{ \Lads^4 - \tilde{z}^4 {\ell^{(0)}}^2 } } }, \\
        z &= \tilde{z} \pqty{1 + \frac{1}{2} \epsilon \qty(f'_+(\tilde{x}^+) + f'_-(\tilde{x}^-)) }.
      \label{lindiffeo2}
\ea

In the new coordinate system,  the metric reads,
\baa
\dd{s}^2 = \Lads^2 \frac{\dd{\tilde{z}}^2}{\tilde{z}^2} + \pqty{\frac{\Lads^2}{\tilde{z}^2} + \frac{\tilde{z}^2}{\Lads^2} \ell_+ (\tilde{x}^+) \ell_- (\tilde{x}^-)   }\dd{\tilde{x}^+} \dd{\tilde{x}^-} + \pqty{ \ell_+ \dd{\tilde{x}^+}^2 + \ell_- \dd{\tilde{x}^-}^2 },  \label{pertban3}
\ea
with
\baa
\ell_\pm (\tilde{x}^\pm) = \ell^{(0)} + \epsilon \pqty{ 2 \ell^{(0)} f'_\pm (\tilde{x}^\pm) - \frac{1}{2} \Lads^2 f'''_{\pm} (\tilde{x}^\pm) }.  \label{pertell2}
\ea

Note importantly that we will be imposing the same boundary condition in this perturbed geometry as in \eqref{affineinit} \emph{in the tilde $(\tilde{.})$ coordinate system}. To wit,
\baa
\eval{\lambda}_{\tilde{z}= \delta} = 0, \qquad  \eval{\dv{\lambda}{\tilde{z}}}_{\tilde{z} = \delta} = \frac{\Lads^2}{\delta^2} - \ell^{(0)}.  \label{affineinit2}
\ea

Let us denote
\baa
g(x^+,  x^-) \equiv \frac{1}{2}  \qty(f'_+(x^+) + f'_-(x^-)).  \label{defgee}
\ea
Now, as mentioned before, in order to make the calculation simpler,  we use the inverse of the coordinate transformation in \eqref{lindiffeo2} and go back to the relatively simple static BTZ metric.  It is the same as the initial unperturbed geometry save the boundary condition. We now have a ``wiggly'' boundary defined from \eqref{lindiffeo2} by
\baa
z = \delta \pqty{1 +\epsilon g(x^+, x^-) },  \label{zeehyper}
\ea
where always working to linear order in $\epsilon$,  we have dropped the tilde in the above defining equation for the boundary hypersurface.  

We will have essentially the same solution for the affine parameter as before \eqref{affinezee},  in terms of the functional dependence on $z$,  but there will be a new additive constant and a new normalization, which we parametrize with two unknown constants $a$ and $b$,
\baa
b\lambda = a- \pqty{ \frac{\Lads^2}{z} + \ell^{(0)}z  }.  \label{affinezee2}
\ea
Imposing the initial condition \eqref{affineinit2} and the equation describing the boundary \eqref{zeehyper}, we determine the constants to be
\begin{subequations}
\begin{align}
a &= \frac{\Lads^2}{\delta}  \pqty{1 -  \epsilon g } + \ell^{(0)} \delta \pqty{1 + \epsilon g }, \label{consdefa}  \\
b &= 1 - \epsilon g \frac{\Lads^2 + \ell^{(0)} \delta^2 }{\Lads^2 - \ell^{(0)} \delta^2}. \label{consdefb}
\end{align}
\label{consdefs}
\end{subequations}
This yields, to order $\epsilon$, the affine parameter at the horizon crossing $z = z_0$,
\baa
\lambda &= \frac{z_0 - \delta}{z_0 \delta} (\Lads^2 - \ell^{(0)} z_0 \delta)  \\
&\quad - \epsilon g \frac{1}{z_0 (\Lads^2 -  \ell^{(0)} \delta^2 )} \bqty{ \Lads^4 + { \ell^{(0)} }^2 z_0^2 \delta^2 + \Lads^2  \ell^{(0)} (z_0^2 - 4z_0 \delta + \delta^2 ) }
.\label{lambdafin}
\ea

We define the key observable $\Delta \lambda (x^+, x^-)$  in the following way:
\baa
\Delta\lambda (x^+, x^-) &\equiv \lim_{\delta \to 0} \pqty{\lambda(\epsilon) - \lambda(\epsilon \to 0) } \\
&= - \epsilon g (x^+, x^-) \pqty{ \frac{\Lads^2}{z_0} + \ell^{(0)}z_0  } \\
&=  - \epsilon g (x^+, x^-) r_0.
\label{deflambda}
\ea
The above is the order of limits prescription under which the final expression is well-defined and simple.  The next task before us is to find correlation functions of the observable $\Delta \lambda$.

\subsection{Calculation of Holographic Correlators}\label{subsec-holocorr}

We are now interested in calculating the two-point holographic correlation function $ \expval{ \Delta\lambda (x^+, x^-) \Delta\lambda (y^+, y^-)  }$ of the observable $\Delta\lambda$ on the boundary theory.  Since both the correlator and the boundary stress tensor depend on the functions $f_\pm$ associated with the diffeomorphism,  we wish to associate the $\Delta\lambda$ correlator with the stress tensor two-point function.

The boundary stress tensor is directly related to the functions $\ell_\pm$ as \cite{Balasubramanian:1999re}
\baa
T_{\pm \pm} (x^\pm) = \frac{1}{8\pi \GN \Lads} \ell_{\pm} (x^\pm) = \frac{1}{12 \pi} \frac{c}{\Lads^2}  \ell_{\pm} (x^\pm) \label{Tpmdef},
\ea
where
\baa
c = \frac{3\Lads}{2\GN} \label{bhcc}
\ea
is the standard Brown-Henneaux central charge \cite{Brown:1986nw}.

For simplicity, let us first  work only with the left-moving sector (+) and turn off the right-movers. (We will duly amend this in the next section)  Decomposing the  stress tensor in a background piece and a perturbation, 
\baa
T _{++} = T^{(0)}_{++} + \delta T_{++} ,  \label{pertpp}
\ea
let us write the correlator of the perturbation in the thermal state as
\baa
 \expval{ \delta T_{++} (x^+)  \delta T_{++} (y^+)  } \equiv S_+ (x^+ - y^+),  \label{pertcorr}
\ea
where we are agnostic about the precise form of the function $S_+$ at the moment. It is convenient to work in the momentum basis
\baa
S (k_+) =  \int \dd{x^+} e^{\mi k_+ x^+} S_+ (x^+).\label{fouriers}
\ea

Let us use $F_+ (x^+) \equiv f'_+(x^+)$ as the fundamental variable.  Then using \eqref{pertell2},  we see that
\baa
 \delta T_{++} (k_+) = \epsilon  \frac{1}{12\pi} \frac{c}{\Lads^2} \pqty{ 2 \ell^{(0)}  + \frac{1}{2} k_+^2 \Lads^2 }  F_+ (k_+).  \label{delttf}
\ea

Therefore, from the definition of the functions,  we can write,
\baa
 \epsilon^2 \expval{ F_+ (k_+) F_+ (p_+) }  = \frac{(12 \pi)^2 \Lads^4}{c^2} 2\pi\delta (k_+ + p_+ ) \frac{ S_+ (k_+) }{ \pqty{ 2 \ell^{(0)}  + \frac{1}{2} k_+^2 \Lads^2 } ^2} .\label{fpsrel1}
\ea

On the other hand,  from the definition of the observable,  \eqref{deflambda} and \eqref{defgee},  we have
\baa
\expval{ \Delta \lambda (k_+) \Delta \lambda (p_+) } = \frac{r_0^2}{4} \epsilon^2  \expval{ F_+ (k_+) F_+ (p_+) }  .\label{delamcor1}
\ea

Thus,  we obtain the simple result,
 \baa
 \expval{ \Delta \lambda (k_+) \Delta \lambda (p_+) } = \frac{(6\pi)^2 r_0^2 \Lads^4}{c^2} 2\pi\delta (k_+ + p_+ ) \frac{ S_+ (k_+) }{ \pqty{ 2 \ell^{(0)}  + \frac{1}{2} k_+^2 \Lads^2 } ^2}. \label{delamcor2}
\ea

This implies that the real-space two-point function,
\baa
\expval{ \Delta \lambda (x^+) \Delta \lambda (0) } =  \frac{18\pi r_0^2 \Lads^4}{c^2} \int \dd{k_+} e^{ -\mi k_+ x^+} \frac{ S_+ (k_+) }{ \pqty{ 2 \ell^{(0)}  + \frac{1}{2} k_+^2 \Lads^2 } ^2}. \label{delamcor3}
\ea

\subsubsection{Wightman Correlator on the Cylinder}

\begin{figure}
\centering
\includegraphics[width=0.4\textwidth]{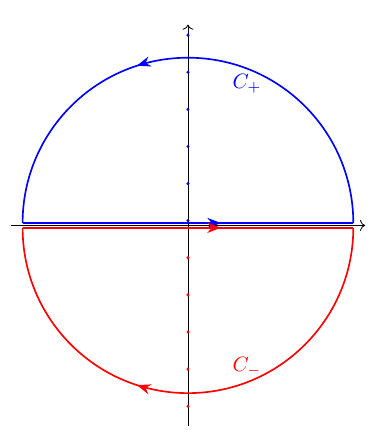}
\caption{The contours used for the evaluation of eq.  \eqref{ftwightde1}}
\label{fig:contours}
\end{figure}

Let us find the Fourier representation $S(k_+)$ by using the Wightman correlator on the cylinder. By using a standard conformal map from the plane to the cylinder, we have,
\baa
S_+ (x^+) = \frac{c}{2}  \pqty{ \frac{\pi}{\beta} }^4 \csch^4 \pqty{ \frac{\pi}{\beta} (x^+ - \mi \epsilon)  }. \label{wightman1}
\ea

We now perform the Fourier transform \eqref{fouriers} in some detail.  We have,
\baa
S_+ (k_+) =  \frac{c}{2}  \pqty{ \frac{\pi}{\beta} }^4 \int_{-\infty}^{\infty} \dd{x^+} e^{ \mi k_+ x^+} \csch^4 \pqty{ \frac{\pi}{\beta} (x^+ - \mi \epsilon)  }.  \label{ftwightde1}
\ea
We evaluate this integral using the Cauchy residue theorem by closing the contour with an infinite semi-circular arc on the upper (lower) half-plane according as $k_+ > 0$ ($k_+ < 0$), as shown in Figure \ref{fig:contours}.

The poles of the integrand are located at $x_+ = \mi (n \beta + \epsilon)$ for $n \in \mathbb{Z}$.  The residue at these poles is given by
\baa
\lim_{\epsilon \to 0} \underset{x_+ = \mi (n \beta + \epsilon)}{\Res} = - \mi \frac{e^{-n k \beta} k_+ \beta^2 (k_+^2 \beta^2 + 4\pi^2 ) }{6\pi^4}.
\ea
The $C_+$ and $C_-$ contours will receive contributions from $n \geq 0$ and $n \leq -1$ respectively.  We can easily sum the geometric series from the residues for each of these contours. Noting that the two contours are oppositely oriented,  we can write down the final answer for the momentum-space stress tensor two-point function,
\baa
 S_+ (k_+ )  = \frac{\pi c}{6} \frac{k_+ \pqty{ k_+^2 + (2\pi/\beta)^2  }}{1 - e^{-k_+ \beta} }. \label{wightmoment}
\ea
The zero-temperature limit $\beta \to \infty$ has the expected form,
\baa
 S_+ (k_+ ) = \frac{\pi c}{6} \theta( k_+) k_+^3 \label{wightmomentzero},
\ea
where $\theta(x)$ is the Heaviside step function.

We can examine various aspects of the momentum-space correlation function \eqref{delamcor2}. We can use the relations \eqref{ellonr} and \eqref{temp1} to write everything in terms of the background temperature,
\baa
\expval{ \Delta \lambda (k_+) \Delta \lambda (p_+)   } =  128 \pi^6 \delta(k_+ + p_+) \GN \Lads^3 T^2 \frac{k_+}{ k_+^2 + 4\pi^2 T^2  } \ \frac{1}{  1- e^{ - k_+ / T }  }.  \label{delamcor5}
\ea
It is worth noting that there has been a cancelation  of one $k^2 + (2\pi T)^2$ factor in the final answer. It is nice to rewrite the answer so that the dependence on the dimensionless quantity $k_+/T$ is transparent,
\baa
\expval{ \Delta \lambda (k_+) \Delta \lambda (p_+)   } =  128 \pi^6 k_+ \delta(k_+ + p_+) \GN \Lads^3 \  \frac{1}{ k_+^2/T^2 + 4\pi^2   }\   \frac{1}{  1- e^{ - k_+ / T }  }.  \label{delamcor5b}
\ea

We can now study the behavior of the correlator in the different regimes,  low and high momenta compared to the temperature,
\baa
\expval{ \Delta \lambda (k_+) \Delta \lambda (p_+)   }  = \delta(k_+ + p_+)   \bqty{ 32 \pi^4 \GN \Lads^3 T + \mathcal{O} (k^+) } .\label{lowkdelam}
\ea
The high momentum behavior is given by
\baa
\expval{ \Delta \lambda (p_+) \Delta \lambda (k_+)   }  = \delta(k_+ + p_+) \frac{128 \pi^6 \GN \Lads^3 T^2}{|k_+|} \times \begin{cases} 1 ,  &\quad k_+ \gg T,  \\
\exp(-|k_+| / T),  &\quad  k_+ \ll - T .
\end{cases}
\label{highkdelam}
\ea

\paragraph{Dependence on horizon size.} We have assumed that the horizon size is large compared to the AdS radius in using our approximation for the stress tensor correlator. Physically, we expect that in this regime, the horizon fluctuations on length scales much shorter than the horizon size are independent of the horizon size. In the above formulae, the black hole radius appears in the temperature $T$; our expectation is that somehow the $T$ dependence should disappear.

Note that the metric near the horizon is 
\be
\dd s^2 \approx  {\dd r^2 \over f} + {r_0^2 \over 4 \Lads^2}(\dd x^+ - \dd x^-)^2,
\ee
so the natural coordinate near the horizon is
\be
\hat x^+ = {r_0 \over 2 \Lads} x^+. \label{xxhatrel}
\ee
Recall that the temperature is $T= r_0/(2 \pi \Lads^2)$ so the relation can be written
\be
\hat x^+ =  \pi \Lads T x^+.
\ee
We can also define a corresponding rescaled momentum conjugate to $\hat x$,
\be
\hat k = {k \over  \pi \Lads  T}.
\ee
It is natural to take the position space affine parameter fluctuation to transform as a scalar,
\be
\hat \lambda(\hat x) = \lambda( x(\hat x)).
\ee
This implies that in Fourier space the variable transforms with an additional factor, as
\be
\hat \lambda(\hat k) =   \pi \Lads T \lambda( \pi \Lads T \hat k).
\ee
In terms of these horizon-adapted variables, we have
\be
\expval{\Delta \hat \lambda(\hat k^+) \Delta \hat \lambda (\hat p^+)} = \hat k_+ \delta(\hat k^+ + \hat p^+) \GN \Lads^5 T^2 \hat{g}(\pi \hat k^+ \Lads),
\ee
where $\hat{g} (x) \equiv 128 \pi^8 \bqty{ (x^2 + 4\pi^2)(1- \exp(-x))}^{-1}$ is the function that appears within \eqref{delamcor5b}.
This expression still carries an overall factor of $T^2$. However, recall that we chose the normalization so that the affine parameter is simply related to the $r$ coordinate, $\dd \lambda= - \dd r$ asymptotically. This arbitrary choice will disappear when we convert to the quantum width, which is defined as a proper distance. This conversion will precisely cancel the factors of $T$, yielding a result independent of the black hole radius. 

Concretely, using \eqref{widthfromdeltar}, we have
\be
\expval{ \widehat{\mathcal{W}}_Q^2(\hat k^+)  \widehat{\mathcal{W}}_Q^2 (\hat p^+)} = \hat k_+ \delta(\hat k^+ + \hat p^+) \GN \Lads^5 {\hat{g}(\pi \hat k^+ \Lads) \over \pi^2},
\ee
which shows that the fluctuations on scales much shorter than the horizon size do not depend on the horizon size, assuming that the black hole radius is much larger than the AdS radius, $r_0 \gg \Lads$.

\subsection{Smeared Correlation Functions and the Quantum Width}\label{subsec-smear}

With the Wightman correlators so defined, we can use smeared correlation functions defined as
\baa
\expval{ \mathcal{O} (f) } = \int_{-\infty}^\infty \dd{x} f(x) \expval{  \mathcal{O}(x) },  \label{smeardef}
\ea
where $f(x)$ is some suitably normalized, rapidly decaying, smooth test functions.  For example,  we can take a Gaussian smearing function
\baa
f_{\rm G} (x) = \frac{1}{ \sqrt{2\pi} \sigma } \exp( -\frac{x^2}{2\sigma^2} ).  \label{gaussian}
\ea
We can also write down a momentum space representation of the correlators using the definition of the Fourier transform in \eqref{fouriers}. For the two-point function, we have for instance,
\baa
\expval{ {\cal O} (f_1) {\cal O} (f_2) } = \frac{1}{(2\pi)^2}\int\int \dd{k} \dd{p} f_1 (-k) f_2 (-p) \expval{ \mathcal{O} (k) {\cal O} (p) }  .\label{smearmo}
\ea
Let us evaluate the smeared two-point function for identical Gaussians \eqref{gaussian}, for which 
\baa
f_G (k) = \exp( - \frac{1}{2} k^2 \sigma^2  ) \label{gaussmo}.
\ea
We thus have,
\baa
 \expval{\Delta \lambda (f_{\rm G})^2 } &= 32\pi^4 \GN \Lads^3 T^2 \int_{-\infty}^\infty \dd{k} \frac{ k e^{ -k^2 \sigma^2 } }{k^2 + (2\pi T)^2}
\frac{1}{  1- e^{ - k / T }}, \label{integral1} \\
&= 32\pi^4 \GN \Lads^3 T^2 I (\sigma T).
\ea
This integral $I$ is purely a function of the dimensionless parameter $\sigma T$.  When $\sigma T \sim 1$,  we can already make an estimate and we find,
\baa
\expval{\Delta \lambda (f_{\rm G})^2 } \sim \GN \Lads^3 T^2. \label{delamest}
\ea

Since we are using a normalization  of the affine parameter so that  $\Delta \lambda \sim \Delta r_0$, see \eqref{affdel2},  can use the relation \eqref{widthfromdeltar} to deduce the quantum width,
\baa
\wq^4 \sim \GN \Lads^3, \label{widthcase2}
\ea
which is quite remarkable in that it is completely independent of the black hole temperature (contrast with \eqref{qwcase1}, for canonical fluctuations) and that it has one power of $\GN$.

After some manipulations, we can write \eqref{integral1} as,
\baa
     \expval{\Delta \lambda (f_{\rm G})^2 } = 16 \pi^4 \GN \Lads^3 T^2 e^{4 \pi^2 \sigma^2 T^2 }  E_1 (4 \pi^2 \sigma^2 T^2) + h(\sigma T) ,  \label{integexp1}
\ea
where $E_1 (z)$ is the exponential integral function \cite{AShandbook} and $h(z )$ is regular for $z \to 0$.  Using the properties of the $E_1$ function for small arguments and the function $h$ for large arguments,  we can find the behavior of the correlator for small and large values of $\sigma T$,
\baa
   \expval{\Delta \lambda (f_{\rm G})^2 } = \begin{cases}
        32 \pi^4 \GN \Lads^3 T^2  |\log (\sigma T)| + \cdots, &\quad \sigma T \ll 1, \\
        \dfrac{8\pi^{5/2} \GN \Lads^3 T}{\sigma} + \cdots, &\quad \sigma T \gg 1. 
    \end{cases} \label{smecollims1}
\ea
There is thus a logarithmic enhancement on the first line of \eqref{smecollims} and a power-law suppression on the second line.

We can translate this to the quantum width, using again \eqref{widthfromdeltar},
\baa
    \expval{(\wq^2 (f_{\rm G}))^2 } = \begin{cases}
        32 \pi^2 \GN \Lads^3  |\log (\sigma T)| + \cdots, &\quad \sigma T \ll 1, \\
        \dfrac{8\pi^{1/2} \GN \Lads^3 }{\sigma T} + \cdots, &\quad \sigma T \gg 1. 
    \end{cases} \label{smecollims}
\ea
It is interesting to relate this to the near-horizon normalized quantities defined above. The smearing width $\sigma$ is related  to a proper length on the horizon via \eqref{xxhatrel}
\begin{equation}
    {\sigma \over 2\Lads} = {\hat \sigma \over r_0}.
\end{equation}
Recalling that $T = {r_0 \over 2 \pi \Lads^2 }$ we have the relation
\begin{equation}
    \sigma T = {\pi \hat \sigma \over \Lads} .
\end{equation}
so that the fluctuations can be written in terms of near-horizon quantities as 
\baa
     \expval{(\wq^2 (f_{\rm G}))^2 } = \begin{cases}
        32 \pi^2 \GN \Lads^3  \left|\log \frac{\hat \sigma}{\Lads} \right| + \cdots, &\quad \hat \sigma  \ll \Lads, \\
        \dfrac{8\pi^{-1/2} \GN \Lads^4 }{\hat \sigma} + \cdots , &\quad \hat \sigma  \gg \Lads. 
    \end{cases}
\ea

We can relate this to the physically transparent heuristic approach adopted between eqs.  \eqref{ncelldef}--\eqref{newscale}. We can write an analogous equation to  \eqref{drtotsq} for the affine parameter,
 \baa
    \overline{\pqty{\Delta \lambda}^2} \sim \frac{1}{N_{\rm cell} } \pqty{\Delta \lambda_{\rm cell} }^2. \label{dlamtotsq}
\ea

We take the $\mathcal{O}(1)$ value \eqref{delamest} for $\pqty{\Delta \lambda_{\rm cell} }^2$ and take $N_{\rm cell} \sim T \sigma\gg 1$,  from which we obtain,
\baa
\overline{\pqty{\Delta \lambda}^2} \sim \frac{\GN \Lads^3 T}{\sigma},  \label{dlamtotsq2}
\ea
This is thus consistent with the second limiting behavior from \eqref{smecollims}.

\section{General Set-Up with a Compact Horizon}\label{sec-various}

In the previous section, we have obtained the central physical result of the paper. In doing so, however, we have made several simplifications for the sake of clarity and concise presentation. In this section, we consider some more general physical situations. The essential lessons we learned in the previous section would still be valid. 

\subsection{Both Left- and Right-Movers Turned On}\label{ssubsec-lrboth}
So far, we have discussed the scenario with only the left-moving sector with the $x^+$ components turned on. However,we must consider the situation in which both sectors are turned on --- since we are talking about the cylinder, this consideration is a necessity.  Many of the formulas described above will go through, with a few small changes.

To begin with, the formulas \eqref{pertpp}--\eqref{fpsrel1} go through in a straightforward manner with replacements $+ \to -$. Now, the two-point momentum space correlator has the following form (as opposed to \eqref{delamcor1}),

\baa
    \expval{ \Delta\lambda (k_+, k_-) \Delta \lambda (p_+, p_- )} = \frac{r_0^2 \epsilon^2}{4} (2\pi)^2 &[\delta(k_-) \delta (p_-) \expval{F_+ (k_+) F_+ (p_+) } \\
    &+ \delta(k_+) \delta (p_+) \expval{F_- (k_-) F_- (p_-) } ].
    \ea
We have already set $\expval{F_+(k_+) F_- (p_-)} = 0 = \expval{F_-(k_-) F_+ (p_+)}$, using the fact that the connected stress-tensor two point function between the left and right movers vanishes for the cylinder. (However, on the torus, there could be possible contact terms and zero momentum terms.) The additional delta functions are consistent with the momentum space dimension of $\expval{ \Delta\lambda (k_+, k_-) \Delta \lambda (p_+, p_- )}$ which is different from before because of two additional Fourier integrals.

Thus, we have,
 \baa
 c^2 \frac{\expval{ \Delta \lambda (k_+, k_-) \Delta \lambda (p_+, p_-) } }{288 \pi^5 r_0^2 \Lads^4 } &= \delta(k_-) \delta(p_-)  \delta (k_+ + p_+ ) \frac{ S_+ (k_+) }{ \pqty{ 2 \ell^{(0)}  + \frac{1}{2} k_+^2 \Lads^2 } ^2} \\
 &\quad +  \delta(k_+) \delta(p_+)  \delta (k_- + p_- ) \frac{ S_- (k_-) }{ \pqty{ 2 \ell^{(0)}  + \frac{1}{2} k_-^2 \Lads^2 } ^2}.
   \ea
Correspondingly, the real space correlator is given by
\baa
\expval{ \Delta \lambda (x^+,x^-) \Delta \lambda (0) } &=  \frac{18\pi r_0^2 \Lads^4}{c^2} \int \dd{k_+} e^{ -\mi k_+ x^+} \frac{ S_+ (k_+) }{ \pqty{ 2 \ell^{(0)}  + \frac{1}{2} k_+^2 \Lads^2 } ^2} \\
&\quad + \frac{18\pi r_0^2 \Lads^4}{c^2} \int \dd{k_-} e^{ -\mi k_- x^-} \frac{ S_-(k_-) }{ \pqty{ 2 \ell^{(0)}  + \frac{1}{2} k_-^2 \Lads^2 } ^2}.  \label{realspace2}
 \ea

Since we have defined $x^\pm = t \pm L \phi$ so that the temporal coordinate has a positive coefficient for both the light-cone coordinates,  we will simply have $S_-(x^-) = S_+ (x^-)$ where $S_+$ is of the form \eqref{wightman1}. Much of the remaining calculations go through straightforwardly. The shift in many cases is additive in nature --- the smeared correlator has the same overall form \eqref{integexp1}, now with another additive term. Therefore, the physics is essentially the same as before, with  two independent smearing directions. There would be enhancement on the correlator even if just one of the directions is considered as before. In the position space, there are independent divergences along $x^+$ and $x^-$, as one would expect in such scenarios.

\subsection{A Horizon of Finite Size}\label{ssubsec-finite}

In the discussion of the Wightman correlation function so far, we have considered the two-point stress tensor correlation function on the cylinder, \eqref{wightman1}. This means that the angular direction $\phi$ is non-compact. Thus, the horizon at $r = r_0$ is a not a true black hole horizon, but rather analogous to a Rindler horizon. It turns into a black hole horizon under the identification $\phi \sim \phi + 2\pi$. In this case, one would analytically continue the torus stress tensor correlator and repeat the previous calculations.

However, for the present investigation, we can make use of the large value of the CFT central charge $c$ and therefore apply the method of images to deduce the corresponding correlation function from the cylinder correlator that respects the periodicity under $\phi \sim \phi + 2\pi$. Recalling  the definition of $x^\pm \equiv  t \pm \Lads \phi$, we can do this easily from eq. \eqref{wightman1} as
\begin{equation}
    \wt{S}_+ (x_+) = \sum_{n_+ \in \mathbb{Z} } S_+ (x_+ + 2\pi n_+ \Lads), \label{wighttilde1}
\end{equation}
where the $n_+$-sum is rapidly convergent because of the nature of the $S_+$ function \eqref{wightman1}. The function $\wt{S}$ is by construction doubly periodic under $x^+ \sim x^+ + 2\pi \Lads$ and $x^+ = x^+ + \mi \beta$. It immediately follows that the Fourier transform of \eqref{wighttilde1} will be given by,
\begin{equation}
    \wt{S}_+ (k_+) = \sum_{n_+ \in \mathbb{Z}} e^{-2\pi \mi n_+ k_+ \Lads} S_+ (k_+), \label{ftperwide1}
\end{equation}
where  \eqref{ftperwide1} is given by \eqref{ftwightde1} whose properties were studied extensively. We can rewrite the above by a more useful distributional identity,
\begin{equation}
    \wt{S}_+ (k_+) = \frac{1}{\Lads} S_+ (k_+) \sum_{n_+ \in \mathbb{Z}} \delta \pqty{ k_+ - \frac{n_+}{\Lads} } .\label{ftperwide2}
\end{equation}
To obtain the analog of the smeared correlator \eqref{integral1}, we simply have to insert the sum of delta functions inside the integral in \eqref{integral1}. In the previous case, the smeared correlator was just a function of the dimensionless ratio $\sigma T$. It will now be a function of two dimensionless parameters $\sigma T$ and $\Lads T$,
\begin{equation}
\begin{aligned}
 \wt{\expval{\Delta \lambda (f_{\rm G})^2 }} &= 32\pi^4 \GN \Lads^3 T^2 \sum_{ n_+ \in \mathbb{Z}} \frac{n_+}{n_+^2 + (2\pi T \Lads )^2}\frac{\exp( - n_+^2 \frac{(\sigma T)^2}{(T \Lads)^2} )}{1 - \exp( - \frac{n_+}{T\Lads} )}  \label{integral2sum} \\
&= 32\pi^4 \GN \Lads^3 T^2 \wt{I} (\sigma T, T \Lads).
\end{aligned}
\end{equation}
Even without considering any closed-form solution, we can deduce important properties of the finite horizon case.

First, we note that when $\sigma T$ is held fixed, but $T \Lads \to \infty$, the regime describes precisely the non-compact case we have considered so far. Therefore, in this limit, we recover the earlier function
\begin{equation}
    \wt{I} (\sigma T \text{ fixed}, T \Lads \to \infty) = I (\sigma T),  \label{ltinftylim}
\end{equation}
where $I(\sigma T)$ is defined in \eqref{integral1}.

The interesting limit where the finite size of the black hole is important is the opposite limit, where $T \Lads \ll 1$. In this case, only the $n = 0$ term in \eqref{integral2sum} contributes and we get a different answer, that is independent of $\sigma$ in the leading order,
\begin{equation}
    \wt{I} (\sigma T \text{ fixed}, T \Lads \ll 1) \approx \frac{1}{4\pi^2 T \Lads}.  \label{ltzerolim}
\end{equation}
This means that in this regime,
\begin{equation}
\begin{aligned}
 \wt{\expval{\Delta \lambda (f_{\rm G})^2 }} &\sim \GN \Lads^2 T , \label{integral2sum2} \\
\wq^4 &\sim \frac{\GN \Lads^2}{T}.
\end{aligned}
\end{equation}
This indicates that the fluctuations become large in this regime. We do not trust the answer from the canonical ensemble in this regime anyway since when $T\Lads \ll 1$, we know that the temperature is below the Hawking-Page transition and the black hole phase is sub-dominant compared to the thermal gas. One must therefore go to a microcanonical description in this energy range to make a meaningful analysis.

Let us perform a more detailed analysis of the $\Delta \lambda$ correlators for different angular momentum modes in the most general situation --- we want to find the correlators of $\Delta \lambda_m (t)$ defined as,
\baa
\Delta \lambda_m (t) = \frac{1}{2\pi} \int_0^{2\pi} \dd{\phi} e^{-\mi m \phi } \Delta \lambda (x^+, x^-). \label{angmommodedef}
\ea

When both the left-movers and right-movers are turned on, eq. \eqref{realspace2} tells us that
\baa
\frac{\expval{ \Delta \lambda (x^+,x^-) \Delta \lambda (y^+, y^-) } }{  32\pi^4 \GN \Lads^3 T^2} &=  \int \dd{k_+}  \frac{ e^{ -\mi k_+ (x^+ - y^+)} k_+ }{ (k_+^2 + 4\pi^2 T^2) (1- e^{-k_+/T} )}  \\
&\quad + \int \dd{k_-}  \frac{ e^{ -\mi k_+ (x^- - y^-)} k_- }{ (k_-^2 + 4\pi^2 T^2) (1- e^{-k_-/T} )}.
\ea

We can get a discrete version of this formula by using \eqref{ftperwide2}. Using this formula taking the Fourier modes of both the $\phi$ coordinates involved, we eventually find
\baa
\expval{ \Delta \lambda_m (t) \Delta \lambda_n (t') } = 32\pi^4 \GN \Lads^3 T^2  \delta_{m +n, 0} \bqty{ e^{ \mi \frac{m}{\Lads} (t-t') } \widetilde{\mathcal{S}}_{-m} + e^{ -\mi \frac{m}{\Lads} (t-t') } \widetilde{\mathcal{S}}_{m} },\label{delamdisccorr}
\ea
where
\baa
\widetilde{\mathcal{S}}_{m}  = \frac{m}{\qty(m^2 + 4\pi^2 \Lads^2 T^2) \qty( 1 - \exp( - \frac{m}{\Lads T}  ) ) }.  \label{tildesmdef}
\ea
Note that while the Wightman correlation functions such as eq. \eqref{delamcor5} discriminate between positive and negative values of the light-cone frequency $k_+$ by construction, when we consider the correlation functions between different angular momentum modes the positive and negative values of the modes are treated completely democratically, as can be seen in eq. \eqref{delamdisccorr} above when considering the equal time correlator.

\subsection{Properties of the Position Space Correlator}\label{ssubsec-position}

We  briefly consider the position-space correlator of the $\Delta \lambda$ observable. For simplicity, let us revert to the non-compact horizon with the only the left-mover,
\begin{equation}
\expval{ \Delta \lambda (x^+) \Delta \lambda (0)   } = \frac{1}{(2\pi)^2} \int_{-\infty}^\infty \int_{-\infty}^\infty \dd{p_+} \dd{k_+} e^{- \mi k_+ x^+ } \expval{ \Delta \lambda (k_+) \Delta \lambda (p_+)   } 
    \label{posspa1}.
\end{equation}
We will use the representation \eqref{delamcor5} and we will decompose the integral in the following manner,
\begin{equation}
  \expval{ \Delta \lambda (x^+) \Delta \lambda (0)   } = I_1 (x^+) + I_2 (x^+) ,\label{posspa1d}
\end{equation}
where,
\baa
    I_1 (x^+) &=  32\pi^4 \GN \Lads^3 T^2 \int_0^\infty \dd{k_+} \frac{e^{-\mi k_+ x^+} k_+ }{k_+^2 + 4\pi^2 T^2}, \label{posspa1d1}\\
    I_2 (x^+) &= 32\pi^4 \GN \Lads^3 T^2 \int_{-\infty}^{\infty} \dd{k_+} \frac{e^{-\mi k_+ x^+} |k_+| }{k_+^2 + 4\pi^2 T^2} \frac{e^{-|k_+|/T} }{1-e^{-|k_+|/T} }.
\ea

It is easy to see that a short distance divergence as $x^+ \to 0$ arises only from the first integral $I_1 (x^+)$ --- the integral diverges logarithmically when $x_+ = 0$. The second integral $I_2 (x^+)$, on the other hand, is finite in this limit. Thus, we have to examine only the first integral. The integral can be represented \cite{AShandbook} after an analytic continuation as
\begin{equation}
    \frac{I_1 (x^+)}{32\pi^4 \GN \Lads^3 T^2} = - \cosh(2\pi T x^+ ) \mathrm{Ci} \qty(2 \pi \mi T x^+) + \mi \sinh(2 \pi T x^+) \pqty{ \frac{\pi}{2} - \mathrm{Si} \qty(2\pi \mi T x^+) }, \label{i1exp}
\end{equation}
where $\mathrm{Ci} (z)$ and $\mathrm{Si} (z)$ are the cosine and sine integral functions respectively, which are closely related to the exponential integral function $E_1 (z)$ in \eqref{integexp1}. The expansion of the $\mathrm{Si}(z)$ function about $z = 0$ is analytic, but we do have,
\begin{equation}
    \mathrm{Ci} (z)  \sim \log z + \mathcal{O} (z^0).  \label{ciasympt}
\end{equation}
Therefore, we confirm the presence of the expected short distance logarithmic singularity directly from this representation.

\section{Finite Cut-Off Surfaces and the Poincar\'e Horizon}\label{sec:limits}

In this section, we examine a couple of physically interesting limits pertaining to the correlator.

\subsection[The $\Delta \lambda$ Correlator with a Finite Cut-Off]{\boldmath The $\Delta \lambda$ Correlator with a Finite Cut-Off}\label{ssubsec-finitecutoff}

We first consider the question as to what happens when we send in the light rays from a finite point in the bulk, not from asymptotic infinity. There are numerous ways to address this question. One concrete way to do this is to consider a constant $z = \delta$ surface in the Fefferman-Graham gauge. Luckily, most of the work is already done in the preceding calculation, before we take the limit $\delta \to 0$. To be concrete,  all the calculations between eqs. \eqref{affinezee} and \eqref{lambdafin} go through completely unchanged -- in these calculations, the expressions were correct up to linear order in $\epsilon$, but were always exact in $\delta$.

Therefore,  we can define the finite-$\delta$ affine parameter observable in pretty much the same way as in \eqref{deflambda}, only without taking the limit $\delta \to 0$,
\baa
    \Delta \lambda_\delta (x^+, x^-) =- \epsilon g \frac{1}{z_0 (\Lads^2 -  \ell^{(0)} \delta^2 )} \bqty{ \Lads^4 + { \ell^{(0)} }^2 z_0^2 \delta^2 + \Lads^2  \ell^{(0)} (z_0^2 - 4z_0 \delta + \delta^2 ) }.\label{defdeltalamb1}
\ea
Inserting the explicit values of $\ell^{(0)}$ and $z^0$ in terms of $r_0$ or $T$, we find the expression,
\baa
     \Delta\lambda_\delta (x^+, x^-) &= -\epsilon g(x^+, x^-)  r_0 \frac{2\Lads^2 - r_0 \delta}{2\Lads^2 + r_0 \delta} \\
     &=\frac{1 - \pi T \delta}{1+ \pi T \delta}  \Delta \lambda (x^+ , x^-) ,  \label{dellamdel2}
\ea
where $\Delta \lambda$ is the quantity considered so far in the limit $\delta \to 0$. Thus, we have a simple multiplicative renormalization of the observable in the finite cut-off case. Every factor of $\Delta\lambda$ in a correlator would receive this correction. As a consistency check,  we see that the multiplicative factor is less than unity and goes to zero in the limit $\delta \to 1/\pi T$, i.e., when the boundary approaches the horizon. It would be interesting to explore this scenario in more detail vis-\`{a}-vis the proposed relation between finite cut-off holography and the $T\overline{T}$-deformation 
\cite{McGough:2016lol}.

\subsection{Poincar\'e Horizon}\label{subsec-poincare}

We are now interested in examining the limit $T \to 0$ or $r_0 \to  0$, in which the metric \eqref{nrbtz} takes the form,
\baa
\dd{s}^2 = - \frac{r^2}{\Lads^2} \dd{t}^2 + \frac{\Lads^2}{r^2} \dd{r}^2 + r^2 \dd{\phi}^2,  \label{limpoincare}
\ea
where now there is the Poincar\'e horizon present at $r = 0$. We could go about it in two ways: 
\begin{enumerate}
\item[a.] We could try do an \textit{ab initio} calculation in the metric \eqref{limpoincare}.
\item[b.] We could take  the $T \to 0$ limit of the preceding formulas involving the exact finite-$T$ results.
\end{enumerate}
Option (a) appears to be complicated, at least at first sight, because the Poincar\'e horizon $z = \infty$ is left fixed by the linearized diffeomorphism \eqref{lindiffeo1}.  From this perspective,  the Poincar\'e horizon appears to be rigid.  However,  the approximation of linearized diffeomorphism breaks down around the Poincar\'e horizon anyway.  The non-linear case is of course much more difficult to deal with.

Therefore,  we go with option (b) of taking the zero-temperature limit.  We see from \eqref{smecollims} that in this  limit $\expval{\Delta \lambda (f_G)^2}$ vanishes in spite of the logarithmic enhancement, because of the more dominant $T^2$ pre-factor.  In this sense,  the Poincar\'e horizon appears to be quite rigid. We however end up with a large log if we consider the quantum width instead,
\baa
\wq^4 \sim \GN \Lads^3 \log( \frac{1}{\sigma T} ).  \label{qwlowtemp}
\ea
This is an interesting behavior of the quantum width which merits a physical explanation. 

Let us go back to the definition of the quantum width, \eqref{defqw1}, where we use the exact red-shift factor $f(r)$ instead of the near-horizon linearization,
\baa
\wq = \int_{r_0}^{r_0 + \Delta r_0} \frac{\dd r\,  \Lads}{\sqrt{r^2 - r_0^2}} = 2\Lads \log( \sqrt{\frac{\Delta r_0}{4\pi \Lads^2 T}} + \sqrt{1+ \frac{\Delta r_0}{4\pi \Lads^2 T}}  ).  \label{wdefexact}
\ea
In the limit $T \ll \Delta r_0/\Lads^2$,  we find that
\baa
\wq \sim \Lads \log( \frac{\Delta r_0}{\Lads^2 T}  ).   \label{qwlowtemp2}
\ea
Thus, we could get a log from the exact $f(r)$. However,  comparison between \eqref{qwlowtemp} and \eqref{qwlowtemp2} shows that the powers of log are different and furthermore,   there is a missing power of $\GN$ outside (which in the earlier case was supplied by $\Delta r_0$ itself).

We now argue that we are never in a regime where eq. \eqref{qwlowtemp2} is applicable, by imposing a suitable cut-off. We can require that the minimum black hole size for which our considerations apply is of the Planck scale, which imposes a lower bound on the temperature
\baa
    T > T_{\rm min} \sim \frac{\ell_{\rm Pl}}{\Lads}. \label{timincut}
\ea
In a related manner, we can also demand that the minimum proper width of the Gaussian smearing function on the black hole horizon be the Planck length, for which,
\baa
    \sigma T \sim \sigma \frac{r_0}{\Lads^2} \sim \frac{\sigma_{\rm proper}}{\Lads}  > \frac{\ell_{\rm Pl}}{\Lads}. \label{sigmacutoff}
\ea
We must have $\ell_{\rm Pl} \ll \Lads$ and in the worst-case scenario when this bound is saturated, the first line of \eqref{smecollims} is applicable and we have,
\baa
  \frac{\wq^2}{\Lads^2} \sim \frac{\Delta r}{\Lads^2 T} \sim \sqrt{ \frac{\ell_{\rm Pl}}{\Lads} \log \frac{\Lads}{\ell_{\rm Pl}} }  \ll 1.  \label{wextrcase}
\ea
Thus, we are always in the regime where the previously used formula for the quantum width is applicable. Furthermore, the quantum width is also small in AdS units even in the limiting situation where the black hole horizon goes over to the Poincar\'e horizon.

\section{Discussion}\label{sec-discuss}
The results obtained in the paper are very satisfactory. We have demonstrated, for the first time, a direct and precise calculation of a parametrically large quantum gravity effect within the purview of perturbative quantum gravity and holography. Our results can be extended and generalized in several ways. In this paper, we have discussed one particular protocol for calculating the quantum width of the black hole horizon. One can conceive several other observational protocols, which might not even require the presence of a horizon.  Let us discuss another operational protocol that could show an enhanced fluctuation.

\subsection{Another Protocol for Boundary Observation}

\begin{figure}
    \centering
    \includegraphics[width=0.5\linewidth]{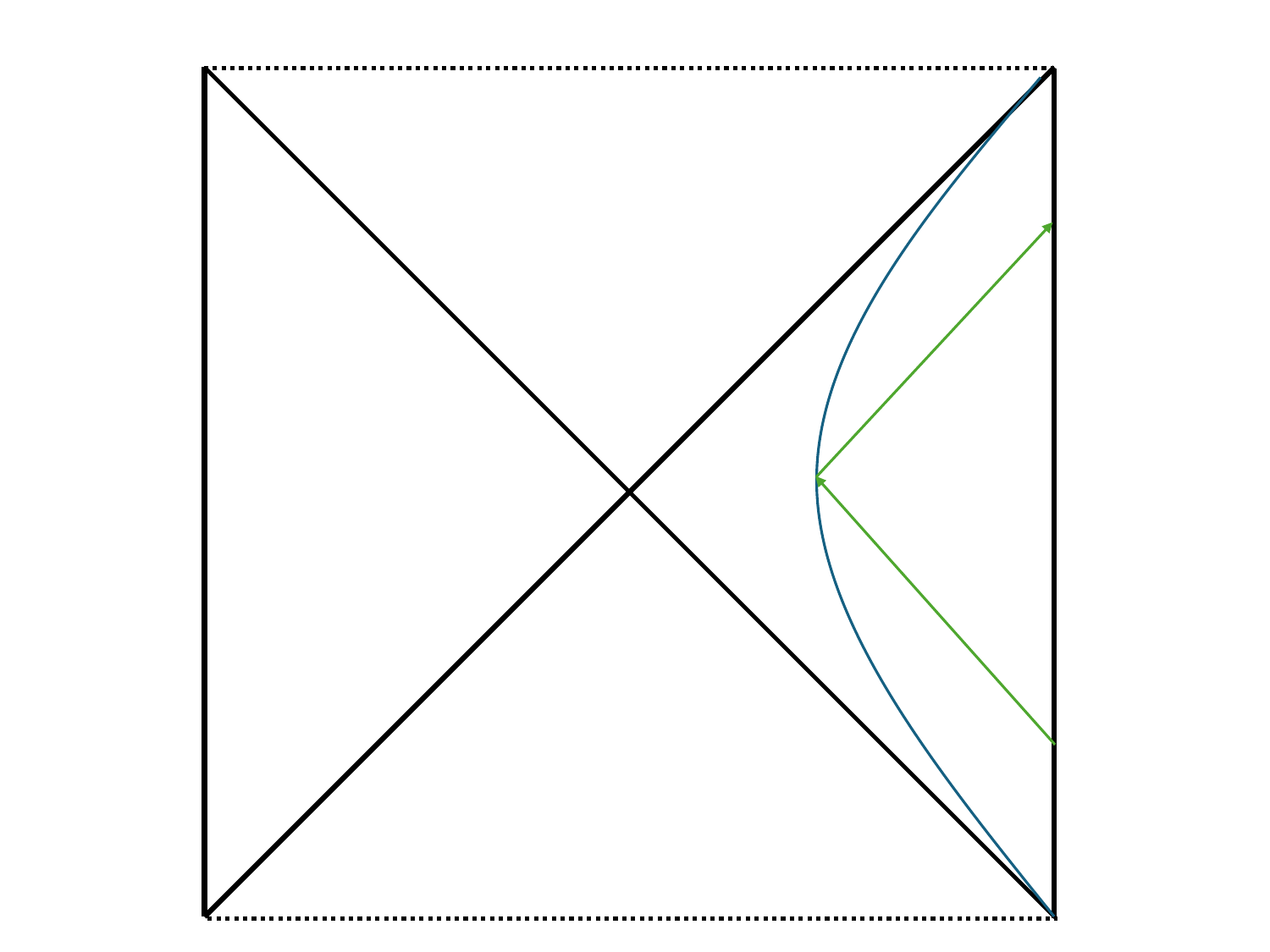}
    \caption{An illustration of the protocol discussed in the text. In blue is the world-like of a mirror at a fixed radial position in the bulk. An observer sends a light ray (in green) towards the bulk and receives the reflection back after some time interval.}
    \label{fig:protocol}
\end{figure}

 An observer on the boundary defined by the coordinate system $\tilde{z} = \delta$ and  $\tilde{x}^\pm = \tilde{x}^\pm_{\rm initial}$ radially sends a light ray towards the bulk. There is a bulk mirror at a fixed position, which is specified  in a gauge-invariant way by fixing the induced metric and extrinsic curvature of the mirror. Since this specification refers to local bulk quantities, the location of the mirror is fixed in the un-tilded coordinates of Fig. \ref{fig:cpdc}, at $r = r_m$.  This reflected null ray hits the boundary at $\tilde{z} = \delta$ at the point $\tilde{x}^\pm = \tilde{x}^\pm_{\rm final}$.  We find the time difference between the initial and final in the un-tilde coordinate. This protocol is depicted in Figure \ref{fig:protocol}.

In the un-tilde coordinate, with a mirror at $r=r_m$, the total time taken for the light ray is finite even as the boundary is sent to infinity,
\begin{equation}
t_{\rm total}= \frac{\Lads^2}{r_0} \log \frac{r_m + r_0}{r_m - r_0}.
\end{equation}

Therefore, in this protocol, where we fix the mirror coordinate in the bulk, the time difference vanishes identically,
\begin{equation}
   \Delta  t_{\rm total} = 0.
\end{equation}

This is, however, not the full story. For a boundary observer,  the tilde coordinate \eqref{lindiffeo2}, is the natural coordinate, we have actually,
\begin{equation}
    \Delta \tilde{t}_{\rm total} = \frac{1}{2}\epsilon \pqty{ f_+ (x^+_{\rm final}) - f_- (x^-_{\rm final}) - f_+ (x^+_{\rm initial}) + f_- (x^-_{\rm initial}) }.
\end{equation}
With no loss of generality, we can choose $x^\pm_{\rm initial} = 0$ and let us call $x^\pm_{\rm final} = x^\pm$. We can then perform a similar Fourier-space analysis as in \S\ref{sec-corrf} and find the correlation function of the $\Delta \tilde{t}$ observable. We find after a calculation, 
\begin{equation}
\expval{\Delta \tilde{t}_{\rm total}(x^+) \Delta \tilde{t}_{\rm total}(y^+)} \sim\frac{\GN}{\Lads} \int \dd{k_+} \frac{\pqty{ e^{-\mi k_+ x^+} -1 } \pqty{ e^{\mi k_+ y^+} -1 } }{k_+ (k_+^2 + 4\pi^2 T^2)(1-e^{-k_+ / T})}. \label{timecorraltpro}
\end{equation}
Notice that even in this expression, there is only one power of the Newton constant $\GN$. Therefore, we see a similar enhancement to the one discussed previously. Note also that the zero-temperature behavior is log-divergent, as we saw for the quantum width as well.

\subsection{Outlook}

With the previous protocol in mind, we emphasize yet again that our calculation does not require the presence of the black hole horizon as an essential component. In fact, for the cylindrical correlator considered -- which can be taken to describe a large black hole -- it is really the Rindler horizon that is being probed here. We can draw more general lessons from our calculations to wider scenarios.

There are several ways in which we can extend our results. We have worked to the leading order in the Newton constant $\GN$. It would be very interesting to explore higher order effects in $\GN$ and the corresponding physics. In particular, we have seen that at leading order smearing in space or in time is enough control UV divergences. This occured because the gravitons are either left- or right-moving, so smearing in space is equivalent to smearing in time. However, once one-loop correction involve the backreaction of matter fields, which have more bulk degrees of freedom. It seems likely that these effects would be more UV-divergent. The perturbative calculation of the quantum width might look something like
\begin{equation}
    \expval{\wq^4} \sim  \GN \Lads^3  \log { \Lads \over \hat \sigma}  + \GN^2 {\Lads^3 \over \epsilon_{\rm UV}} \ \ \ \ \ (?), 
\end{equation}
where $\epsilon_{\rm UV}$ refers to an ultraviolet length scale set by the resolution of the experiment, which may not be tied to the spatial smearing $\hat \sigma$. Because of the special properties of gravitational perturbations in 3d, it may be that these loop corrections dominate over the tree-level calculation we have performed here.

Since the black holes we consider are in asymptotically AdS spacetime, one could even use the full machinery of AdS/CFT to calculate nonperturbative corrections. One would need to understand the nonperturbative definition of the observable.

In addition, we have studied the static BTZ black hole with a radially infalling null probe. We can of course consider similar calculations with more exotic geometries and probes.  Related work on higher dimensions will appear soon \cite{FSV2026}.

It would be worthwhile to refine aspects of our calculation. We used the method of images for the calculation pertaining to a finite-size horizon. However, to be more precise, we should take into account holographic torus correlators \cite{He:2023hoj}, where subtleties such as contact terms could be important. We have studied the canonical correlators --- it will be very instructive to study the  correlation functions in the microcanonical ensemble, especially for the low-lying states of the spectrum. A study of higher-point correlators would also be quite illuminating. An important task would be to understand the operators such as $\Delta \lambda$ on the CFT side from a more fundamental point of view.

We hope to address such questions in future work.

\acknowledgments We would like to thank Suzanne Bintanja, Antony Speranza, and  Erik Verlinde for helpful discussions. BF is partially supported by Heising-Simons Foundation
‘Observational Signatures of Quantum Gravity’ QuRIOS collaboration grant.
UM's research is supported by the European Union’s
Horizon 2020 Research and Innovation Programme
(Grant Agreement No. 101115511).

\bibliographystyle{JHEP}
\bibliography{refs.bib}

\end{document}